\documentclass[conference]{IEEEtran}
\IEEEoverridecommandlockouts
\ifCLASSINFOpdf
\else
\fi
\usepackage{xfrac}
\usepackage{amsmath}
\usepackage{multirow}
\usepackage{color}
\usepackage{amssymb}
\usepackage{cite}

\begin{document}
\bstctlcite{IEEEexample:BSTcontrol}

\title{Wireless Powered Cooperative Relaying Systems with Non-orthogonal Multiple Access}
%
\author{\IEEEauthorblockN{
Ferdi KARA}
\IEEEauthorblockA{
Wireless Communication Technologies Laboratory (WCTLab) \\
Department of Electrical and Electronics Engineering\\
Zonguldak Bulent Ecevit University\\
Zonguldak, TURKEY 67100\\
Email: f.kara@beun.edu.tr}
}

%
%

\markboth{International Conference on Communications}%
{Kara \MakeLowercase{and} Kaya: Spatial Multiple Access (SMA): Enhancing performances of MIMO-NOMA systems}
%
\maketitle
\begin{abstract}
Non-orthogonal multiple access (NOMA) and the cooperative relaying systems are two of the promising techniques to meet requirements of future wireless networks such as high spectral efficiency and wide coverage area. On the other hand, the energy efficiency has also high priority in the applications with limited energy such as sensor networks and/or internet of things (IoTs). To this end, in this paper, we propose wireless powered cooperative relaying system with  NOMA thereby increasing spectral and energy efficiency. We consider three different energy harvesting (EH) protocols (i.e., power sharing (PS), time sharing (TS) and ideal) and for all three EH protocols, we derive achievable rate for the considered system model. We validate the analysis with computer simulations and present the effectiveness of wireless powered system compared to the benchmark.
\end{abstract}
%
\IEEEpeerreviewmaketitle
\section{Introduction}
%
%
%
%
With the exponential increase in connected-devices to the internet, such as smart phones, tablets, watches etc. \cite{VNI}, the future wireless networks are to have ability to meet massive-type communication in ultra-dense networks. Besides, it is also expected to have almost $\% 100$ coverage area and provide very-high spectral efficiency \cite{8805289}. In order to meet these requirements, the interplay between cooperative communication and non-orthogonal multiple access (NOMA) has attracted recent attention from researches\cite{Kim2015a,Jiao2017,Xu2016,Zhang2018,Kara2020,Abbasi2019,Wan2019,Kara2019,8746442,Li2019}.

The decay in spectral efficiency of cooperative communication has been resolved by NOMA implementation and it is proved that NOMA-based cooperative relaying systems (NOMA-CRS) provide higher spectral efficiency \cite{Kim2015a}. Thus, NOMA-CRS have been analyzed widely in terms of achievable rate, outage probability and bit error rate. The ergodic rate of NOMA-CRS have been analyzed over Rayleigh \cite{Kim2015a} and Rician \cite{Jiao2017} fading channels. Then, a novel receiver design has been proposed for NOMA-CRS in \cite{Xu2016}. NOMA-CRS with two different transmission strategies have been investigated in \cite{Zhang2018}. Ergodic rate and outage probability are derived with channel estimation errors. The bit error probability of NOMA-CRS over Nakagami-m fading channels has been derived in \cite{Kara2020} and optimum power allocation is proposed to achieve minimum bit error rate. Moreover, NOMA-CRS with an amplify-forward relay has been analyzed in terms of capacity in\cite{Abbasi2019}. NOMA-CRS with multiple relays have been also analyzed. When two relays are available in the system, NOMA-based diamond relaying systems have been analyzed over Rayleigh fading channels in terms of achievable rate \cite{Wan2019} and bit error probability\cite{Kara2019}. Relay selection schemes for NOMA-CRS have been also investigated and in order to achieve maximum sum-rate in NOMA-CRS, two stage relay selection algorithms have been proposed \cite{8746442}. Furthermore, NOMA-CRS with spatial modulation have been analyzed in terms of achievable rate and bit error rate \cite{Li2019}. However, all aforementioned studies consider that the relay in NOMA-CRS has independent power source and energy harvesting has not been considered.

On the other hand, the energy consumption is also very crucial for future wireless networks. Especially in energy-limited applications such as sensor networks and IoT applications, energy harvesting (EH) from radio waves gains credit. To this end, wireless power transfer (WPT) along with information transferring have taken tremendous attention \cite{6489506,7010878}. Simultaneous wireless information and power transfer (SWIPT)  have been analyzed in both cooperative \cite{Babaei2018} and NOMA networks\cite{Tang2019,Diamantoulakis2016,Hedayati2018,7445146}. Thus, an energy efficient communication system can be implemented. In EH networks, the relay harvests energy from source-to-relay radio link in order to re-transmit data to the destination. However, as mentioned above, to the best of the authors' knowledge, none of the literature researches considers EH in NOMA-CRS.

In this paper, we propose NOMA-CRS with WPT. We investigate achievable rate performance of NOMA-CRS with three different EH protocols (i.e., power sharing (PS), time sharing (TS) and ideal protocols) and derive ergodic rate of NOMA-CRS over Rayleigh fading channels. The analysis is validated via computer simulations and the achieved performance gain is presented compared to the benchmark -NOMA-CRS without WPT/EH-.

The rest of paper is organized as follows. In section II, the NOMA-CRS with WPT is introduced and the EH for all three protocols are defined. In section III, the achievable rate for NOMA-CRS with WPT is derived.  Then, the analytical analysis is validated  via computer simulations in Section IV. The comparisons with the benchmark -NOMA-CRS without WPT/EH- are also presented in this section. Finally, section V discusses the results and concludes the paper.

\section{System Model}
As shown in Fig. 1, a cooperative relaying system is considered where a source (S) is willing to reach out the destination (D) and a decode-forward relay (R) is helping for it. In order to increase spectral efficiency, NOMA is implemented in the first phase and the relay can harvest its energy from RF signal in the first phase to transmit signal in the second phase.\footnote{The required energy to detect and/or modulate signal at the relay is neglected and all the harvested energy is used to transmit signals.} We assume all nodes are equipped with single antenna. The flat fading channel coefficient ($h_k$, $k=sr,sd,rd$) between each node follows $CN(0,\sigma_k^2)$ where $\sigma_k^2$ denotes the large-scale fading coefficient (i.e., path-loss driven by the distance between nodes). $\gamma_k=\left|h_k\right|^2, \ k=sr,sd,rd$ is defined. \begin{figure}[!t]
   \centering
     \includegraphics[width=8cm,height=3cm]{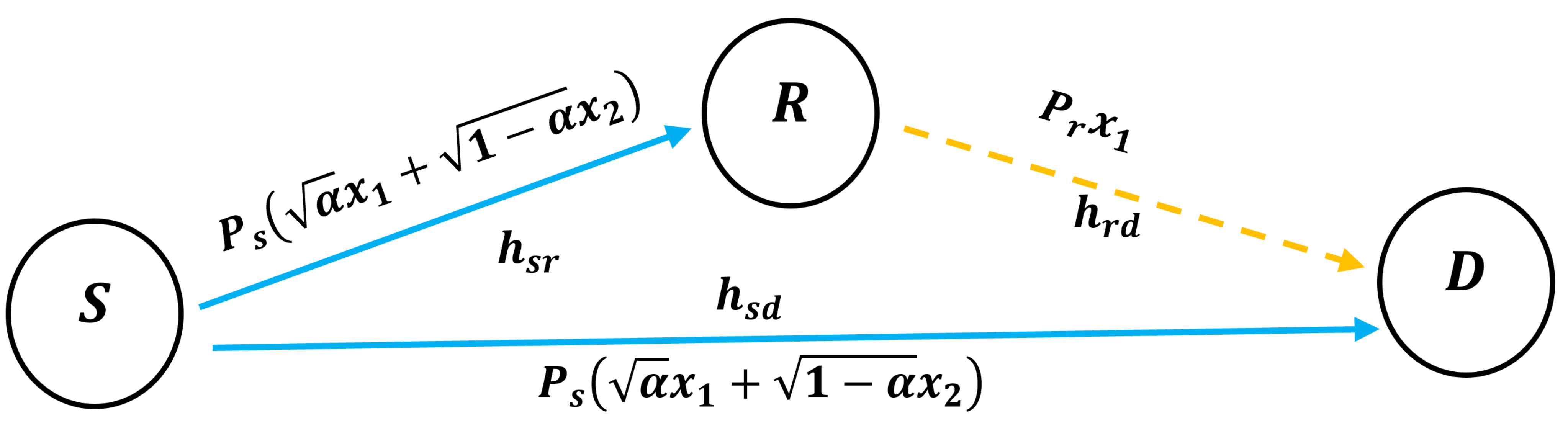}
    \caption{The considered cooperative relaying system with NOMA }
    \label{fig1}
 \end{figure}
  \begin{figure*}[!t]
   \centering
     \includegraphics[width=17cm,height=3.5cm]{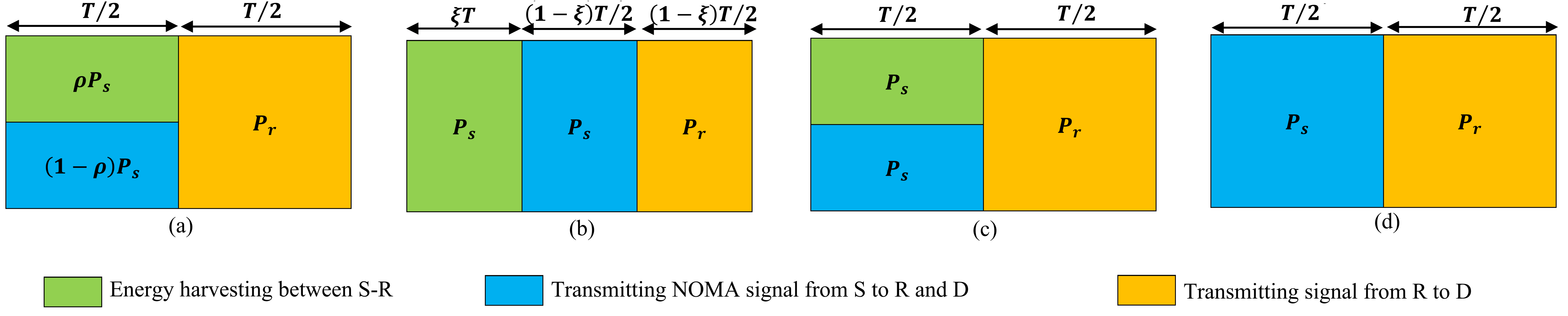}
    \caption{Time schedules for a) EH with PS protocol b) EH with TS protocol c) EH with ideal protocol d)Benchmark (No EH) }
    \label{fig2}
 \end{figure*}
As shown in Fig. 2, three different EH protocols are implemented which are called power sharing (PS), time sharing (TS) and the ideal protocols. In Fig 2, we also present the benchmark where no energy harvesting is available. In the first phase, the source implements superposition coding for two symbols of destination ($\sqrt{\alpha}x_1+\sqrt{\left(1-\alpha\right)}x_2$) where $\alpha$ is the power allocation coefficient (i.e., $\alpha<0.5$) and transmits it with a source power according to implemented EH protocol. Thus, the received signals at the relay and destination are given as
\begin{equation}
 y_k=\sqrt{pP_s}\left(\sqrt{\alpha}x_1+\sqrt{\left(1-\alpha\right)}x_2\right)h_k+n_k, \ k=sr,sd,
 \end{equation}
where $p$ changes according to implemented EH protocol and $p=1$ in TS and ideal protocols whereas $p=1-\rho$ in PS protocol where $\rho$ is PS factor. $n_k$ is additive Gaussian noise and follows $CN(0,N_0)$.

Then, in the second phase, the relay implements successive interference canceler (SIC) and forwards $x_1$ symbols to the destination. Thus, the received signal at the destination is given as
\begin{equation}
        y_{rd}=\sqrt{P_r}x_1h_{rd}+n_{rd},
 \end{equation}
 where $P_r$ is the transmit power of the relay which is harvested from the source-relay link in the first phase.
 \subsection{Source and Relay Powers According to EH Protocol}
 \subsubsection{PS Protocol}
 As seen from Fig. 2(a), in the PS protocol, the source consumes energy during half of the total system duration. Let assume, the total power $P_t$ is consumed during $T$ seconds, hence, the source power in PS mode is given by $P_s=2P_t$. The harvested energy in PS mode is defined
 \begin{equation}
       EH=\eta\rho P_s \left|h_{sr}\right|^2 (\sfrac{T}{2}),
 \end{equation}
 where $0<\eta<1$ is energy conversion coefficient. This harvested energy EH is consumed by the relay within the remained $\sfrac{T}{2}$ time interval. Thus, the relay power in PS mode is obtained as $P_r=\eta\rho P_s\gamma_{sr}$.
\subsubsection{TS Protocol}
In TS protocol, the source consumes energy during $\sfrac{(1+\xi)T}{2}$. Thus, by considering the total power consumption, the source power in TS mode is obtained as $P_s=\sfrac{2P_t}{(1+\xi)}$. The harvested energy at the relay during the first $\xi T$ time interval is given
 \begin{equation}
       EH=\eta P_s \gamma_{sr} (\xi T),
 \end{equation}
and this harvested energy is consumed by the relay within $\sfrac{(1-\xi)T}{2}$ time interval. Hence, the relay power is given by $P_r=2\eta(\sfrac{\xi}{(1-\xi)})P_s\gamma_{sr}$.
\subsubsection{Ideal Protocol}
In the ideal protocol, the source transmits power in the first $\sfrac{T}{2}$ time interval and the relay harvests it. Following the steps above and below of (3) and/or (4). The source power in the ideal protocol is $P_s=2P_t$. The harvested energy
 \begin{equation}
       EH=\eta P_s \gamma_{sr} (\sfrac{T}{2})
 \end{equation}
and the relay power is $P_r=\eta P_s\gamma_{sr}$.
\subsection{Benchmark (No WPT/EH)}
If the relay has not able to harvest energy, the total power is shared between the relay and the source. Since both the source and the relay transmission cover half of total $T$ time interval, the powers of the source and the relay are given by $P_s=P_r=P_t$.
\section{Achievable Rate Analysis}
In order to derive achievable rate, we should firstly define signal-to-interference plus noise ratios (SINRs) between each node for both symbols.

According to (1), since the $x_2$ symbols have more power, the relay and the destination detects them by pretending $x_1$ symbols as noise. Thus, the received SINRs for $x_2$ symbols between S-R and S-D are given as
\begin{equation}
\begin{split}
    &SINR_{x2}^{(sr)}=\frac{(1-\alpha)pP_s\gamma_{sr}}{\alpha pP_s\gamma_{sr}+N_0}, \\
     &SINR_{x2}^{(sd)}=\frac{(1-\alpha)pP_s\gamma_{sd}}{\alpha pP_s\gamma_{sd}+N_0}.
\end{split}
\end{equation}
We hereby remind that $p$ changes according to used EH protocol and the $P_s$ is the source power defined for that protocol in the previous section.
At the relay SIC is implemented to obtain $x_1$ symbols, thus the received SINR for $x_1$ symbols between S-R is given
\begin{equation}
    SINR_{x1}^{(sr)}=\frac{\alpha p P_s\gamma_{sr}}{N_0}.
\end{equation}
Lastly, according to (2), the received SINR for $x_1$ symbols between R-D is given as
\begin{equation}
    SINR_{x1}^{(rd)}=\frac{P_r\gamma_{rd}}{N_0}.
\end{equation}
We again note that $P_r$ differs for each protocol as defined in the previous section.

The achievable rate in a cooperative relaying system is limited by the weakest link. Thus, the achievable rates for the symbols are given by
\begin{equation}
\begin{split}
    & R_1=\min\{R_1^{(sr)},R_1^{(rd)}\}, \\
    &R_2=\min\{R_2^{(sr)},R_2^{(sd)}\}.
\end{split}
\end{equation}
By using Shannon rate formula \cite{ShannonC1956}, the achievable rates are obtained as
\begin{equation}
\begin{split}
    & R_1=\zeta\log_2\left(1+\min\{SINR_1^{(sr)},SINR_1^{(rd)}\}\right), \\
    &R_2=\zeta\log_2\left(1+\min\{SINR_2^{(sr)},SINR_2^{(sd)}\}\right),
\end{split}
\end{equation}
where $\zeta$ coefficient exists since only a definite/part of time interval is used for transmission through S-R/D or R-D. Thus, $\zeta$ differs according to EH protocol and it is given as
\begin{equation}
\zeta\triangleq
\begin{cases}
\frac{1}{2}, & \text{in PS and ideal protocols}, \\
\frac{1-\xi}{2}, & \text{in TS protocol}.
\end{cases}
\end{equation}

In order to obtain ergodic rates, we firstly substitute (6), (7) and (8) into (10) and then average over instantaneous $\gamma_k$.  Let analyze $x_2$ symbols firstly, with some algebraic manipulations, the ergodic rate (capacity) of $x_2$ symbols is given by
\begin{equation}
\begin{split}
   C_2=&\int\limits_0^\infty{\zeta\log_2\left(1+Yp\sfrac{P_s}{N_0}\right)f_Y(y)dy}\\
   &-\int\limits_0^\infty{\zeta\log_2\left(1+Y\alpha p\sfrac{P_s}{N_0}\right)f_Y(y)dy},
\end{split}
\end{equation}
where $Y\triangleq\min\{\gamma_{sr},\gamma_{sd}\}$. $f_Y(y)$ is the probability density function (PDF) of $Y$ and it is given by
\begin{equation}
f_Y(y)=\left(\frac{1}{\sigma^2_{sr}}+\frac{1}{\sigma^2_{sd}}\right)\exp\left(-y\left(\frac{1}{\sigma^2_{sr}}+\frac{1}{\sigma^2_{sd}}\right)\right), y\geq0.
\end{equation}

And the ergodic rate for $x_1$ symbols are given by (14) (see the top of the next page).
\begin{figure*}[t]
\centering
\begin{equation}
\begin{split}
 C_1=\int\limits_0^\infty\int\limits_0^\infty\zeta\log_2\left(1+\min\{\sfrac{\alpha pP_s}{N_0}\gamma_{sr},\sfrac{ P_r}{N_0}\gamma_{rd}\}\right)
 f_{\gamma_{sr}}(\gamma_{sr})f_{\gamma_{rd}}(\gamma_{rd})d\gamma_{sr}d\gamma_{rd}
\end{split}
\end{equation}
\hrulefill
\end{figure*}

It is noteworthy that $P_r$ in (14) changes according to EH protocol. Hence, by substituting $P_r$ derived in Section II.A for each EH protocol and with some simplifications, the erdogic rate of $x_1$ symbols turns out to be
\begin{equation}
 C_1=\int\limits_0^\infty\zeta\log_2\left(1+\sfrac{P_s}{N_0}Z\right)f_Z(z)dz,
\end{equation}
where we define $Z\triangleq\gamma_{sr}W$ and $W\triangleq\min\{p\alpha,\Upsilon\gamma_{rd}\}$. According to EH protocol,
\begin{equation}
\Upsilon=\begin{cases}
\eta\rho, & \text{in PS protocol}, \\
2\eta\frac{\xi}{1-\xi}, & \text{in TS protocol}, \\
\eta, & \text{in ideal protocol} \\
\end{cases}
\end{equation}
is defined.

In order to derive PDF of $Z$, we firstly derive PDF of $W$. Since the $\alpha p$ is a constant and $\gamma_{rd}$ is exponentially distributed, the PDF W is obtained as
\begin{equation}
\begin{split}
f_W(w)&=exp\left(-\frac{\alpha p}{\Upsilon\sigma^2_{rd}}\right)\delta\left(w-\alpha p\right)\\
&+\frac{1}{\Upsilon\sigma^2_{rd}}exp\left(-\frac{w}{\Upsilon\sigma^2_{rd}}\right)\left(1-u\left(w-\alpha p\right)\right), \ \  w\geq0,
\end{split}
\end{equation}
where $\delta()$ and $u()$ denote unit impulse and unit step functions, respectively.
The PDF of Z is obtained as
\begin{equation}
\begin{split}
f_Z(z)&=\int\limits_0^\infty f_w(w)f_{\gamma_{sr}}(\sfrac{z}{w})\frac{1}{w}dw \\
&=\frac{1}{\alpha p\sigma^2_{sr}}\exp\left(-\frac{z}{\alpha p\sigma^2_{sr}}-\frac{\alpha p}{\Upsilon \sigma^2_{rd}}\right) \\&+\int\limits_0^{\alpha p} \frac{1}{\Upsilon\sigma^2_{rd}\sigma^2_{sr}w}\exp\left(-\frac{z}{\sigma^2_{sr}w}-\frac{w}{\Upsilon \sigma^2_{rd}}\right)dw.
\end{split}
\end{equation}
By substituting (18) into (15), the ergodic rate of $x_1$ symbols is obtained. Finally, the sum-rate (throughput) of NOMA-CRS with WPT is obtained as
\begin{equation}
    C_{sum}=C_1+C_2,
\end{equation}

\section{Numerical Results}
In this section, we present Monte Carlo simulations to validate theoretical analysis. In all figures, simulations are presented by markers and theoretical curves\footnote{In numerical integration, the infinity in the upper bounds of the integrals is changed with $10^3$ not to cause numerical calculation errors.} are denoted by lines. All simulation results are presented for $10^5$ channel realizations.

In Fig. 3 and Fig. 4, we present ergodic sum-rate of NOMA-CRS with WPT for two different channel conditions those are $\sigma^2_{sr}=3dB$, $\sigma^2_{sd}=0dB$, $\sigma^2_{rd}=3dB$ and  $\sigma^2_{sr}=10dB$, $\sigma^2_{sd}=3dB$, $\sigma^2_{rd}=10dB$, respectively. In both figures, two different power allocations for NOMA are chosen as $\alpha=0.1$ and $\alpha=0.2$. The achievable rate of NOMA-CRS with WPT is presented for all EH protocols (i.e., PS, TS and ideal protocols). In PS protocol, $\rho=0.1,0.3$ and in TS protocol $\xi=0.1,0.2$ are chosen. In all figures, the energy conversion coefficient $\eta=0.95$ is assumed. First of all, it can easily be seen that analytical analysis matches perfectly with simulations. As expected, when the channel conditions are better (i.e., Fig. 4), the achievable rates for all EH protocols are improved.  In all scenarios, PS protocol outperforms TS protocol and NOMA-CRS without EH. Besides, its performance is very close to ideal EH protocol. On the other hand, TS protocol is superior to the case without EH in low SNR region. However, it provides lower sum-rate in high SNR region.
\begin{figure}[!t]
   \centering
     \includegraphics[width=9cm,height=5.5cm]{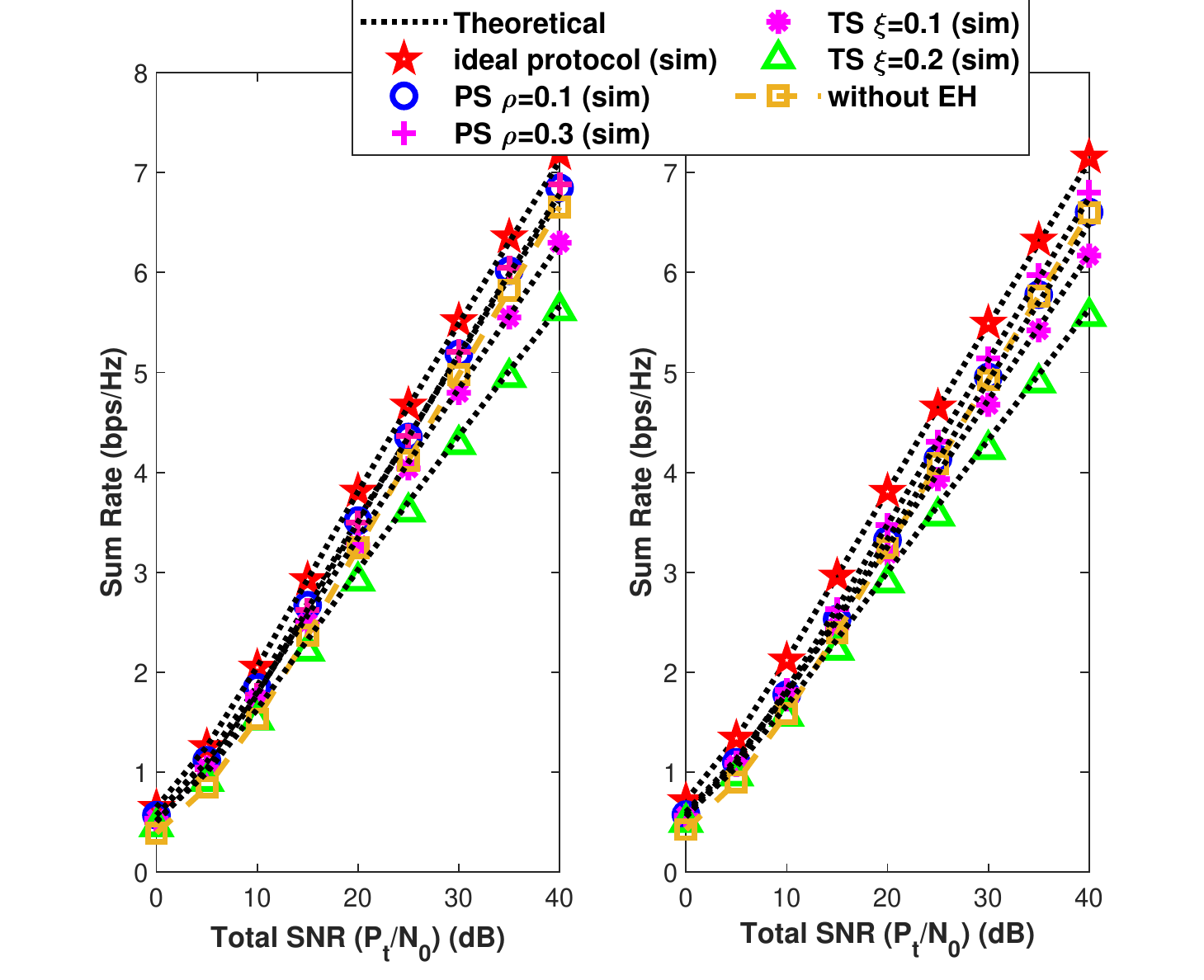}
    \caption{Sum rate of NOMA-CRS with WPT vs. total SNR ($\sfrac{P_t}{N_0}$) when $\sigma_{sr}^2= 3B$,$\sigma_{sd}^2= 0B$ and $\sigma_{rd}^2= 3B$ a) $\alpha=0.1$ b) $\alpha=0.2$}
    \label{fig3}
 \end{figure}
  \begin{figure}[!t]
   \centering
     \includegraphics[width=9cm,height=5.5cm]{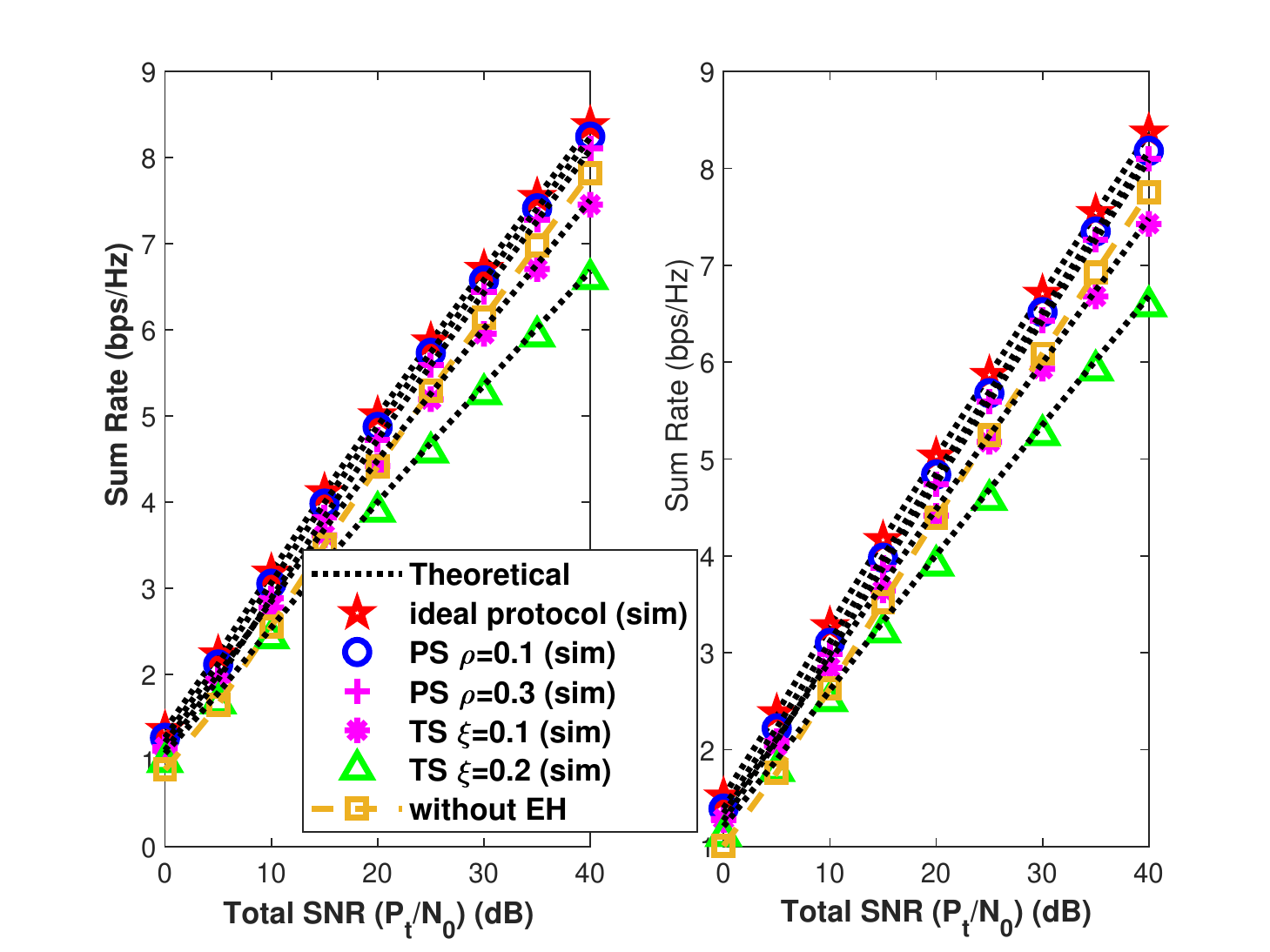}
    \caption{Sum rate of NOMA-CRS with WPT vs. total SNR ($\sfrac{P_t}{N_0}$) when $\sigma_{sr}^2= 10B$,$\sigma_{sd}^2= 3B$ and $\sigma_{rd}^2= 10B$ a) $\alpha=0.1$ b) $\alpha=0.2$}
    \label{fig4}
 \end{figure}

One can easily see from the above-mentioned figures that the chosen parameters for EH (i.e., $\rho,\ \xi$) and the power allocation coefficient (i.e, $\alpha$) have dominant effects on the sum-rate of NOMA-CRS with WPT. To this end, we present sum-rate of NOMA-CRS with WPT with the change of $\rho$ in PS and of $\xi$ in TS protocols in Fig. 5 and Fig 6, respectively. In both figures, the channel conditions are assumed to be $\sigma^2_{sr}=10dB$, $\sigma^2_{sd}=3dB$, $\sigma^2_{rd}=10dB$ and the total transmit SNR (i.e., $\sfrac{P_t}{N_0}$) is fixed to $20dB$. It is clearly seen that according to chosen parameters, both PS and TS protocols may be either close to ideal protocol or worse than the case without EH. The optimum values which provide maximum sum rate for this scenario can be chosen as $\rho=0.135$ in Ps and $\xi=0.025$ in TS protocols. Lastly, in order to reveal the effect of power allocation coefficient on sum-rate, we present the sum-rate with respect to power allocation coefficient (i.e., $\alpha$) in Fig. 7 for the same channel conditions in Fig. 5 and Fig. 6. The EH parameters are chosen according to previous discussions as $\rho=0.135$ and $\xi=0.025$. Based on Fig. 7, under maximum sum-rate constraint and by considering all EH protocols, the optimum power allocation in NOMA-CRS with WPT is seen as $\alpha=0.078$.
\begin{figure}[!t]
   \centering
     \includegraphics[width=9cm,height=5.5cm]{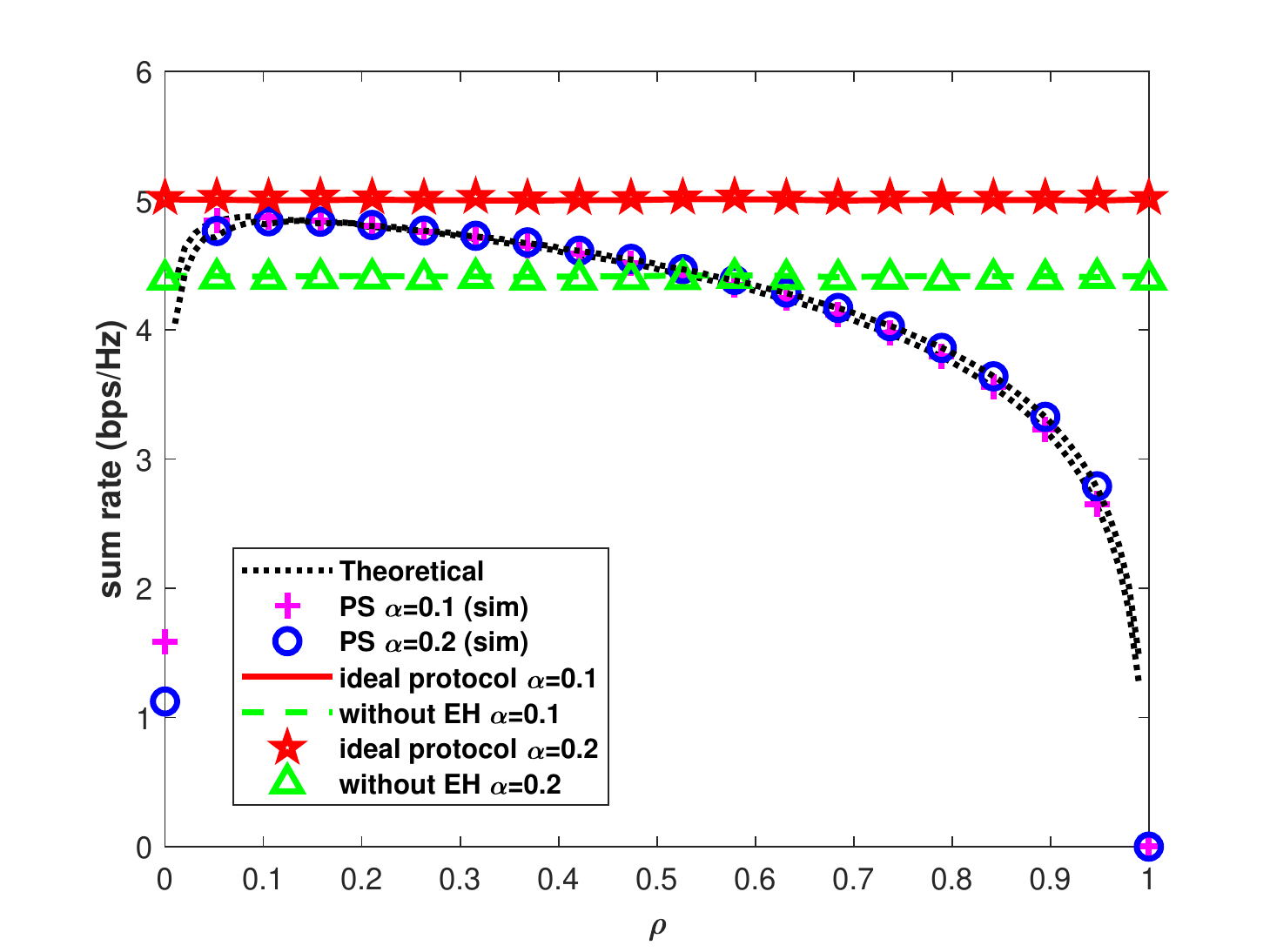}
    \caption{Sum rate of NOMA-CRS with WPT in PS protocol vs. $\rho$  when $\sigma_{sr}^2= 10B$,$\sigma_{sd}^2= 3B$, $\sigma_{rd}^2= 10B$ and $\sfrac{P_t}{N_0}=20dB$}
    \label{fig5}
 \end{figure}
 \begin{figure}[!t]
   \centering
     \includegraphics[width=9cm,height=5.5cm]{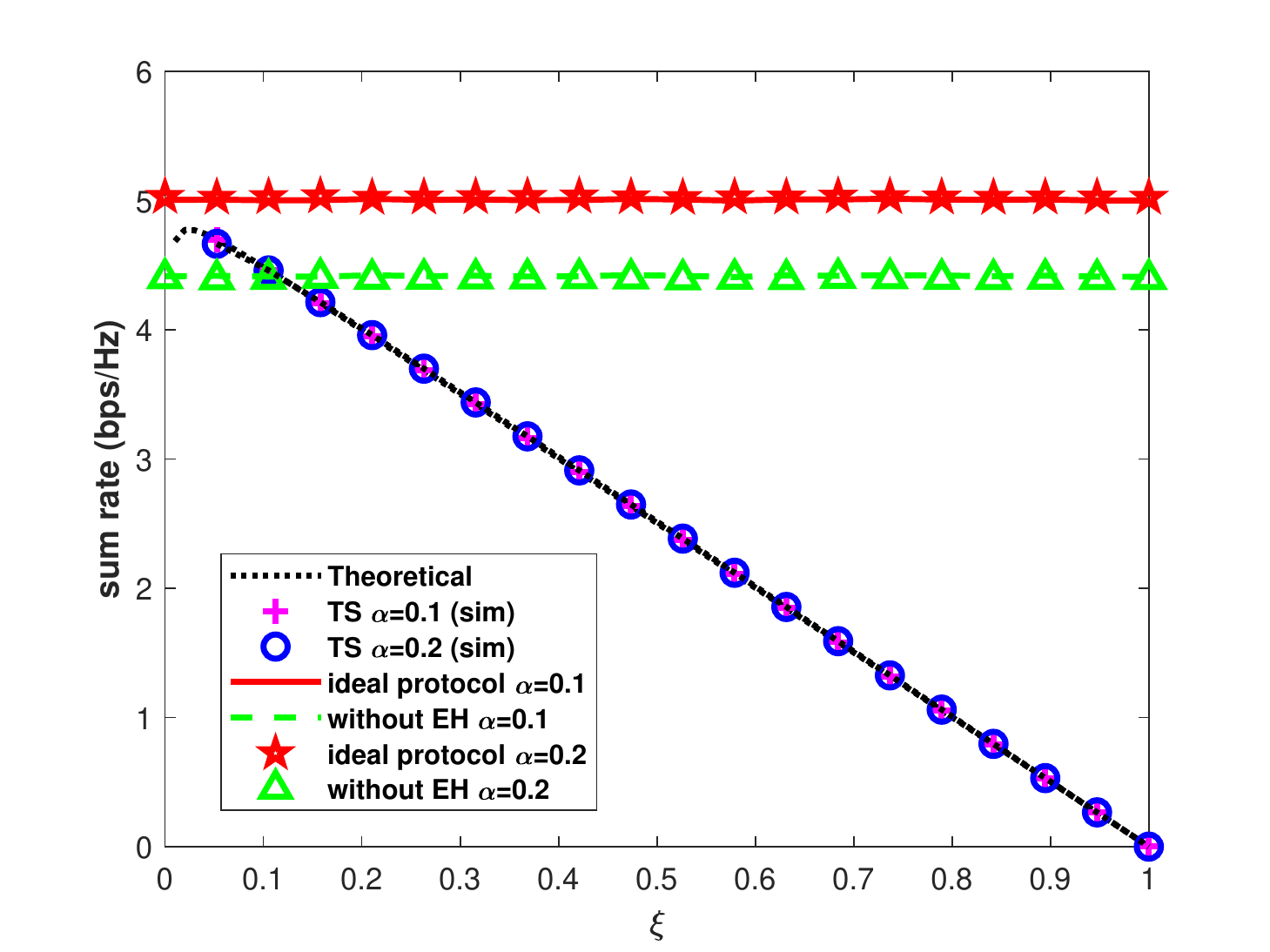}
    \caption{Sum rate of NOMA-CRS with WPT in TS protocol vs. $\xi$  when $\sigma_{sr}^2= 10B$,$\sigma_{sd}^2= 3B$, $\sigma_{rd}^2= 10B$ and $\sfrac{P_t}{N_0}=20dB$}
    \label{fig6}
 \end{figure}
 \begin{figure}[!t]
   \centering
     \includegraphics[width=9cm,height=5.5cm]{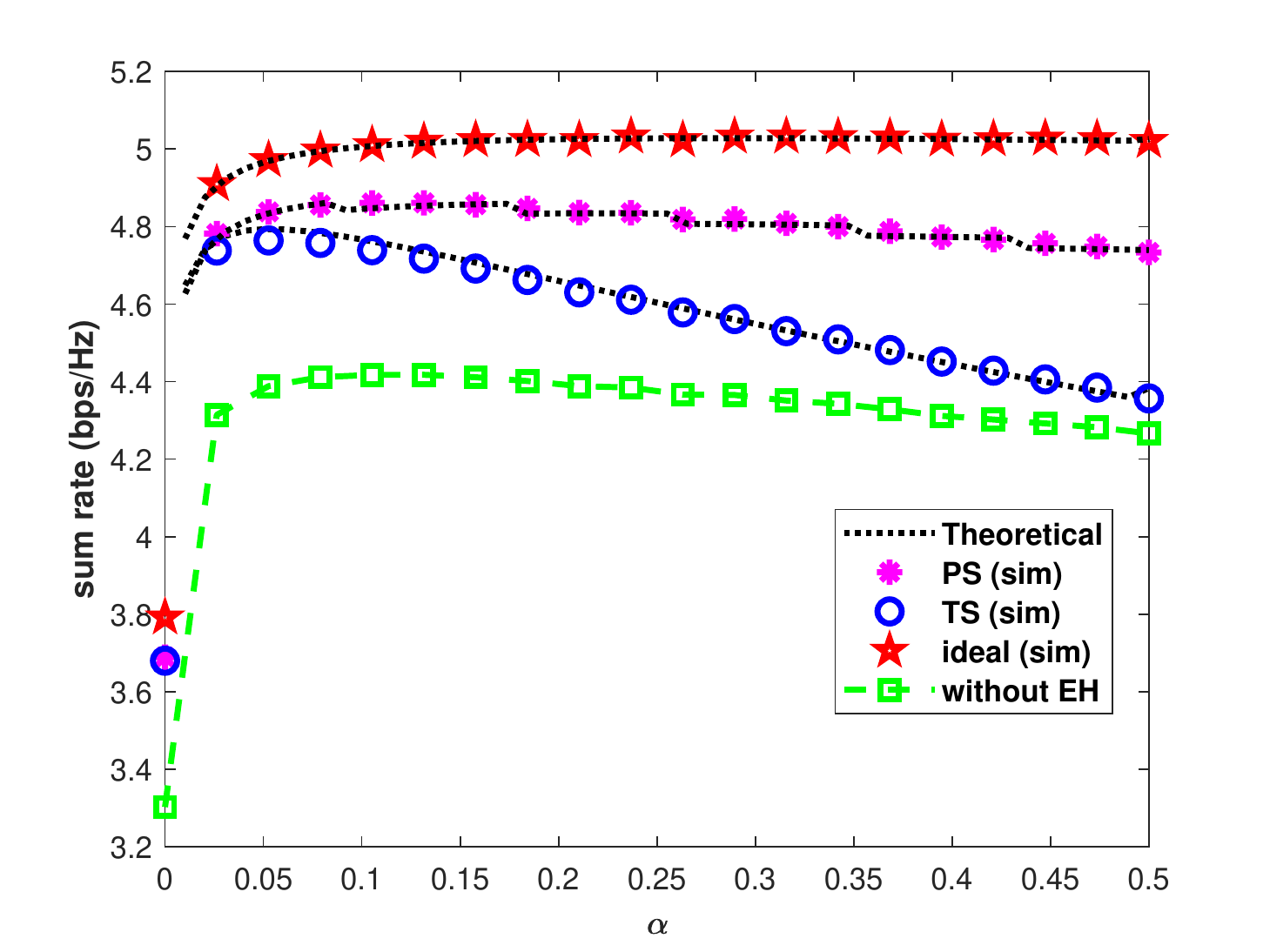}
    \caption{Sum rate of NOMA-CRS with WPT vs. $\alpha$  when $\sigma_{sr}^2= 10B$,$\sigma_{sd}^2= 3B$, $\sigma_{rd}^2= 10B$ and $\sfrac{P_t}{N_0}=20dB$}
    \label{fig7}
 \end{figure}

\section{Conclusion}
In this paper, we propose EH in NOMA-CRS networks. Three different EH protocols (i.e., PS, TS and ideal) are considered  and for all protocols, the achievable rate of NOMA-CRS with WPT is analyzed. The analysis is validated via computer simulations. Based on simulation results, it is revealed that NOMA-CRS with WPT outperforms the benchmark (NOMA-CRS without EH). In other words, higher sum-rate is achieved with the same total energy consumption. Considering the energy efficiency constraint, WPT is seen as promising solution and the WPT implementation in other NOMA-involved systems are the future directions of our researches.

%
\ifCLASSOPTIONcaptionsoff
  \newpage
\fi
%
\bibliographystyle{IEEEtran}
\bibliography{noma_crs_swipt}
%
\end{document}